\documentclass[final,5p,times,twocolumn,authoryear]{elsarticle}

\usepackage{amssymb}

\usepackage[utf8]{inputenc}
\usepackage{adjustbox}
\usepackage{hyperref} 
\usepackage{comment}
\usepackage{graphicx}
\usepackage{xcolor} %

\usepackage{tabularx}
\usepackage{multirow}
\usepackage{textcomp}

\journal{Journal of Systems and Software}

\begin{document}

\begin{frontmatter}

\title{Microservice API Evolution in Practice: A Study on Strategies and Challenges}

\author[inst1,inst2]{Alexander Lercher}
\author[inst1]{Johann Glock}
\author[inst1]{Christian Macho}
\author[inst1]{Martin Pinzger}

\affiliation[inst1]{organization={Department of Informatics Systems, University of Klagenfurt},%
            addressline={Universitätsstraße~65-67}, 
            city={Klagenfurt},
            postcode={9020}, 
            country={Austria, },
            email={\{firstname.lastname\}@aau.at}
            }

\affiliation[inst2]{Corresponding author}

\begin{abstract}
Nowadays, many companies design and develop their software systems as a set of loosely coupled microservices that communicate via their Application Programming Interfaces (APIs). While the loose coupling improves maintainability, scalability, and fault tolerance, it poses new challenges to the API evolution process. %
Related works identified communication and integration as major API evolution challenges but did not provide the underlying reasons and research directions to mitigate them.
In this paper, we aim to identify microservice API evolution strategies and challenges in practice and gain a broader perspective of their relationships.
We conducted $17$ semi-structured interviews with developers, architects, and managers in $11$ companies and analyzed the interviews with open coding used in grounded theory. 
In total, we identified six strategies and six challenges for REpresentational State Transfer (REST) and event-driven communication via message brokers. 
The strategies mainly focus on API backward compatibility, versioning, and close collaboration between teams. The challenges include change impact analysis efforts, ineffective communication of changes, and consumer reliance on outdated versions, leading to API design degradation.
We defined two important problems in microservice API evolution resulting from the challenges and their coping strategies: tight organizational coupling and consumer lock-in.
To mitigate these two problems, we propose automating the change impact analysis and investigating effective communication of changes as open research directions.
\end{abstract}

\begin{keyword}
Microservice architecture \sep API evolution \sep API versioning \sep backward compatibility \sep API design degradation \sep development team collaboration
\end{keyword}

\end{frontmatter}

\newcommand{\todo}[1]{{\color{red} TODO #1}}
\newcommand{\discussion}[1]{\todo{discussion: #1}}

\section{Introduction}

\newcommand{\rqOne}[0]{What means do developers use to exchange messages between services, and how do they document them?}
\newcommand{\rqTwo}[0]{Which strategies do developers follow to introduce and communicate API changes in loosely coupled systems?}
\newcommand{\rqThree}[0]{Which challenges do developers face when introducing and communicating API changes in loosely coupled systems?}

Many modern software systems are split into loosely coupled services to improve maintainability, scalability, and fault tolerance~\citep{9109514}.
Service-oriented Architecture (SOA) \citep{10.5555/1096077} was one method to distribute a large monolithic software system into multiple smaller services. 
SOA relied on shared business models and centralized communication over an Enterprise Service Bus (ESB)~\citep{10.1145/3183628.3183631}. Consequently, the individual services were tightly coupled, and introducing changes required integration and deployment coordination throughout the system's services~\citep{app11177856}.
This led companies to migrate from SOA to Microservice Architecture (MSA). 
MSA replaced the shared models of services with independent domain models exposed only via Application Programming Interfaces (APIs) for each so-called microservice, and the ESB with message brokers only forwarding serialized messages~\citep{8703917}. 
Microservices exposing their functionality via APIs are called \textit{providers} and microservices calling and interacting with these APIs are called \textit{consumers}. 
The communication approach in MSA is referred to as "smart endpoints and dumb pipes" and loosely couples the microservices via well-defined APIs, allowing them to evolve independently within a system~\citep{9825762}.

However, maintaining the overall systems' functionality requires more synchronization efforts between the development teams as each microservice's data structures and business logic, i.e., behaviors, evolve independently~\citep{MA2019724}.
Unlike a single monolithic code base or shared SOA interfaces, the loose coupling prohibits developers from learning about API changes at compile time.
If provider teams do not notify consumer teams about changes in advance, the breaking changes only manifest during the first actual API call at runtime.
Changes in the provider's API could then result in unexpected behavior and potentially break the execution of dependent consumers interacting with that API.

Previous studies~\citep{7796008, app12115507} identified communication and integration as major challenges in the MSA. Similarly, \cite{10.1145/3183628.3183631} identified communicating the API changes to dependent teams, i.e., consumers, and testing for incompatibilities as the primary two open research challenges in service integration.
According to \cite{8938118}, the API evolution process still misses effective communication and support for consumers affected by changes.
\cite{ASSUNCAO2023111788} found that many developers waste their time with implementing technical API changes and updates instead of focusing on business logic.
Hence, microservice API evolution requires more research~\citep{10.1145/3470133}.
While related works acknowledged the challenges of API evolution, integration, and communication of changes, they did not provide the underlying reasons or how to solve them sustainably. 

In this paper, we aim to understand current microservice API evolution strategies and challenges in practice and to gain a broader perspective of their relationships for future research directions. 
For this, we defined the following three research questions:
\begin{itemize}
    \item[RQ1] \rqOne
    \item[RQ2] \rqTwo
    \item[RQ3] \rqThree
\end{itemize}

For answering the three questions, we conducted semi-structured interviews~\citep{doi:https://doi.org/10.1002/9781119171386.ch19} 
with practitioners from multiple companies, analyzed the interviews with open coding~\citep{Corbin1990}, and grounded our qualitative research results with related literature.
Through this, we identified
a) REpresentational State Transfer and event-based communication as the main communication techniques in MSA,
b) six API evolution strategies formulated as best practices for practitioners, 
c) six API evolution challenges to consider as pitfalls when designing an MSA,
d) two important problems namely tight organizational coupling and consumer lock-in, 
and e) two directions for future research to address these problems and improve microservice API evolution.
We provide a replication package~\citep{lercher_alexander_2023_8275799} comprising the interview guide and resulting code book.

To the best of our knowledge, this is the first study investigating microservice API evolution strategies and challenges in practice to create a comprehensive list of best practices and pitfalls and derive two important problems and open research directions to mitigate them.
We focus on loosely coupled services based on MSA and use the terms \textit{service} and \textit{microservice} interchangeably. We intentionally avoid the term \textit{web API} in our work because this often includes Simple Object Access Protocol (SOAP) APIs used in SOA. Instead, we use \textit{API} when referring to microservice communication interfaces. 

The remainder of this paper is structured as follows. 
Section~\ref{sec:meth} describes the method of semi-structured interviews, our study design, data analysis, and participant selection.
Section~\ref{sec:res1} presents the message exchange techniques used in practice and answers RQ1. 
In Sections~\ref{sec:res2} and~\ref{sec:res3}, we present the identified evolution strategies and challenges in practice and answer RQ2 and RQ3.
We discuss the findings, draw the big picture, and define two important problems in API evolution in Section~\ref{sec:discussion}.
Section~\ref{sec:relwork} presents related works on API evolution and Section~\ref{sec:conclusion} concludes our study.

\newcommand{\dto}[0]{object}
\newcommand{\citeinterview}[2]{{\it "#1"}~\mbox{#2}.}
\newcommand{\citeinterviewinline}[3]{{\it "#1"}~\mbox{#2}#3}

\newcommand{\showstats}[4]{(min=#1; max=#2; mean=#3; median=#4)}

\section{Methodology}
\label{sec:meth}
In this section, we describe our study design, data analysis, and participant selection.

\subsection{Study design} %
\label{sec:meth:design}
Due to the open-ended and explorative nature of our research questions,
we conducted semi-structured interviews~\citep{doi:https://doi.org/10.1002/9781119171386.ch19, Gudkova2018} %
with developers, architects, and project managers directly working on MSA or similar loosely coupled systems.
This approach allows the participants to express their thoughts freely while maintaining the desired dialogue direction. 
We formulated an interview guide focusing on answering our research questions to serve as an orientation during the interviews.

The interview guide consisted of five question categories: a) background,
b) communication, targeting RQ1, c) API evolution as provider, targeting RQ2 and RQ3, d) API evolution as consumer, targeting RQ2 and RQ3 from a different perspective explained below, and e) additional thoughts. 
 
The background category elicits the participants' education, experience, subjective definition of a microservice, and details about their work environments. These questions help to set the context and clarify the terminology used by the interviewer and participant during each interview.
 
The communication category aims at answering RQ1. It focuses on the communication approaches, which and how microservice APIs are exposed, and how they are documented. 

The two API evolution categories explicitly illuminate the provider and consumer sides in API evolution to answer RQ2 and RQ3.
During preliminary discussions, we realized that developers do not think about the evolution of external APIs but expect unlimited availability of the consumed API version. 
Hence, we decided to explicitly split the perspectives on provided APIs, i.e., APIs developed and maintained by the interview participants' teams, and consumed APIs, i.e., APIs interacted with by the participants' teams without access to the source code or runtime environment. 
The categories contain open questions regarding the frequency of provided and consumed APIs' changes, the reasons for these changes, the strategies for communicating and implementing the changes, the strategies for notifying other teams about changes, the strategies for getting informed on changes, the challenges they encountered during each of these tasks, and general improvement ideas.

Finally, the interview guide concludes with the question "Do you have additional thoughts you want to express?" sometimes triggering multiple more minutes of dialogue. We used this question to encourage the participants to discuss additional topics we did not consider but that they think are important.

We followed established guidelines for qualitative research~\citep{doi:https://doi.org/10.1002/9781119171386.ch19, doi:https://doi.org/10.1002/9781119171386.ch22} and refined the interview guide two times. After an initial pilot interview, we moved the questions for general improvement ideas from the last category into the provider and consumer API change categories to improve the interview flow. After the fourth interview, we added a background question about the used development and deployment technologies to have a clear picture of the participants' systems. 

In total, we conducted $20$ interviews but excluded three from the results (cf. Section~\ref{sec:meth:participants}).
We designed the interview guide to last between $60$-$90$ minutes. Depending on the available time frame and involvement of the participants, the interviews lasted around 71 minutes \showstats{52}{92}{70.9}{71 minutes}. Due to the SARS-CoV-2 pandemic and physical distance, we conducted $11$ interviews via online videoconference and $6$ interviews in person. We did not observe noticeable differences in the openness or involvement of the participants between these two modes.

\subsection{Interview analysis}
We analyzed the interviews qualitatively to answer the open-ended research questions. 
First, we recorded each interview and transcribed it verbatim.
Then, we applied open coding~\citep{Corbin1990} used in grounded theory~\citep{glaser1967discovery, Adolph2011}.
In this method, individual interview statements, e.g., {\it "You already have a set of test cases that you can run against the old interface. You will see immediately when you introduce something that breaks it"}, are labeled with matching codes, e.g., test the interface on changes. The codes are then assigned to categories, e.g., contract testing, which themselves form a hierarchy, e.g., contract testing is a subcategory of the regression testing strategy.

The first author analyzed all interview transcripts statement by statement, identified the codes, and organized them into a hierarchy of categories. 
We applied investigator triangulation~\citep{doi:10.1177/0049124113500475}, i.e., the second and third authors analyzed two random interview transcripts independently and we discussed the identified categories to increase the result quality~\citep{doi:10.1177/1609406919899220}.
We achieved coder agreement after short discussions, mainly on the phrasing of categories with the same meaning.
We analyzed all interviews iteratively, i.e., using the resulting codebook from the previous session for the next interview transcript. After $12$ analyzed interviews, the codebook began to stabilize, i.e., we only found a few new categories for the following two interviews and the last three interviews did not add any new categories but instead only repeated existing ones. Hence, we reached theoretical saturation~\citep{10.1371/journal.pone.0181689}. 

We structured the categories into the following topics: background, communication and documentation, API evolution strategies, API evolution challenges, and improvement ideas. 
This structure allowed us to answer the research questions directly from the code book. Additionally, we used the findings to build an overall theory of the relationships between strategies and challenges.
In this paper, we used the format ($i$/17) to indicate the number $i$ of participants supporting a finding.

We applied member checking~\cite{doi:https://doi.org/10.1002/9781118181034.ch5} by sharing the study results with our interview participants for feedback and validation. Therefore, we created a draft report and per participant highlighted all findings and statements where we considered their answers. We sent out the $17$ individually highlighted reports and received $13$ responses. Two participants had minor remarks which we incorporated and the others fully agreed with our interpretations. 

Finally, we grounded our findings with related works per category. This approach helped to support or reject our qualitative results and strengthened the overall theory.

\subsection{Participant selection}
\label{sec:meth:participants}

\newcommand{\cOne}[0]{C1}
\newcommand{\oneOne}[0]{C1-P1} %

\newcommand{\cTwo}[0]{C2}
\newcommand{\twoOne}[0]{C2-P1} %
\newcommand{\twoTwo}[0]{C2-P2} %
\newcommand{\twoThree}[0]{C2-P3} %

\newcommand{\cThree}[0]{C3}
\newcommand{\threeOne}[0]{C3-P1} %
\newcommand{\threeTwo}[0]{C3-P2} %
\newcommand{\threeThree}[0]{C3-P3} %

\newcommand{\cFour}[0]{C4}
\newcommand{\fourOne}[0]{C4-P1} %

\newcommand{\cFive}[0]{C5}
\newcommand{\fiveOne}[0]{C5-P1} %
\newcommand{\fiveTwo}[0]{C5-P2} %

\newcommand{\cSix}[0]{C6}
\newcommand{\sixOne}[0]{C6-P1} %

\newcommand{\cSeven}[0]{C7}
\newcommand{\sevenOne}[0]{C7-P1} %
\newcommand{\sevenTwo}[0]{C7-P2} %

\newcommand{\cEight}[0]{C8}
\newcommand{\eightOne}[0]{C8-P1} %

\newcommand{\cNine}[0]{C9}
\newcommand{\nineOne}[0]{C9-P1} %

\newcommand{\cTen}[0]{C10}
\newcommand{\tenOne}[0]{C10-P1} %

\newcommand{\cEleven}[0]{C11}
\newcommand{\elevenOne}[0]{C11-P1} %
\newcommand{\elevenTwo}[0]{C11-P2} %
\newcommand{\elevenThree}[0]{C11-P3} %

\newcommand{\cTwelve}[0]{C12}
\newcommand{\twelveOne}[0]{C12-P1} %

Our participants had to be developers, architects, or managers 
working on developing loosely coupled services exposing an API, e.g., Representational State Transfer (REST) or event-driven communication, for at least one year.
Similarly to other studies~\citep{10.1145/3338906.3338979, 10.1145/3368089.3409743}, 
we contacted previous colleagues and applied snowball sampling~\citep{doi:10.1177/004912418101000205}, i.e., asked them to forward our interview request to their peers matching our requirements as potential participants. 
Considering the explorative nature of the study, this sampling technique is sufficiently effective for theoretical saturation~\citep{Baltes2022}.

We continuously advertised our call for interview participants to colleagues while conducting and analyzing the scheduled interviews.
In total, we contacted $25$ colleagues directly and stopped sending out additional requests once our codebook reached saturation. We only accepted a maximum of three interview partners per company on a first-come, first-served basis. 
Through the snowball sampling, we conducted $20$ interviews with participants from $12$ companies but excluded three of the interviews from the results. One participant worked in a team of only two developers, who created their API solely for the front end and, hence, handled API evolution like any other internal source code change. Another participant did not introduce breaking changes to their product's APIs yet and did not consume any external APIs. The third excluded participant learned of our intermediate results and was excluded to avoid biased answers. 

In total, we report on the results of $n$=$17$ interviews from $11$ companies.
All participants are industry practitioners with an average of 10 years of practical experience \showstats{2}{25}{10.2}{10 years} 
and an average of 4.5 years of practical experience with loosely coupled services \showstats{1}{7}{4.6}{5 years}. Their highest relevant education ranges from a technical high school diploma to a doctoral degree (Ph.D.). 
The technical roles include developers, architects, technical leads, a department head, and a product manager.
Table~\ref{tab:background} contains the details about the individual participants.

\newcommand{\yrs}[0]{}

\begin{table*}[ht]
  \centering
  \caption{%
   Backgrounds of the companies and interviewed participants. 
   \textsuperscript{1}Total practical experience. 
   \textsuperscript{2}Practical experience with loosely coupled services.
   }
  \label{tab:background}

\begin{adjustbox}{width={\textwidth},totalheight={\textheight},keepaspectratio}%
\begin{tabular}{llllllrr}
  \hline 
  Company code & Industry field & Size & Participant code & Highest education & Technical role & Exp\textsuperscript{1} (yrs) & Exp\textsuperscript{2} (yrs) \\
  \hline
  \hline
  \cOne{} & Construction & Large 
  & \oneOne{} & Bachelor & Developer & 7 \yrs{} & 3 \yrs{} \\
   
  \hline
  \multirow{3}{*}{\cTwo} & \multirow{3}{*}{Access management} & \multirow{3}{*}{Large} 
  & \twoOne{} & Ph.D. & Principal architect & 13 \yrs{} & 6 \yrs{} \\
  &&& \twoTwo{} & Ph.D. & Architect & 10 \yrs{} & 6 \yrs{} \\
  &&& \twoThree{} & Master & Architect & 10 \yrs{} & 6 \yrs{} \\
  
  \hline
  \multirow{3}{*}{\cThree} & \multirow{3}{*}{Automotive} & \multirow{3}{*}{Large} 
  & \threeOne   & Master                      & Architect  &  10 \yrs{} &  3 \yrs{}    \\
  &&& \threeTwo   & Bachelor  & Developer & 4 \yrs{} & 4 \yrs{}     \\ %
  &&& \threeThree & Technical high school      & Developer  & 7 \yrs{} & 5 \yrs{}    \\
  
  \hline
  \cFour{} & & & \fourOne{} \textit{(excluded)} &   \\

  \hline
  \multirow{2}{*}{\cFive} & \multirow{2}{*}{Video processing} & \multirow{2}{*}{Medium-sized} 
  & \fiveOne    & Master   & Technical lead / Senior developer & 10 \yrs{} & 4 \yrs{}       \\ %
  &&& \fiveTwo    & Technical high school       & Senior developer  & 7 \yrs{}    & 6 \yrs{}      \\

  \hline
  \cSix{} & Retail & Large 
  & \sixOne{} & Bachelor  & Senior developer / Technical lead  & 15 \yrs{} & 3 \yrs{}  \\

  \hline
  \multirow{2}{*}{\cSeven} & \multirow{2}{*}{Monitoring} & \multirow{2}{*}{Large} 
  & \sevenOne   & Master      & Developer   &     4 \yrs{}   & 3 \yrs{}               \\  %
  &&& \sevenTwo{} \textit{(excluded)}    &              \\ 

  \hline
  \cEight{} & Process digitization & Small 
  & \eightOne   & Bachelor & Developer     & 6 \yrs{}    & 3 \yrs{}    \\

  \hline
  \cNine{} & E-commerce & Small 
  & \nineOne    & Ph.D.         & Developer  & 2 \yrs{}  & 1 \yrs{}       \\ %
  
  \hline
  \cTen{} & Traffic management & Large 
  & \tenOne     & Master      & Architect & 14 \yrs{}   & 7 \yrs{}     \\
  
  \hline
  \multirow{3}{*}{\cEleven} & \multirow{3}{*}{Research and higher education} & \multirow{3}{*}{Large} 
  & \elevenOne  & Master         & Architect  & 9 \yrs{}     & 7 \yrs{}       \\
  &&& \elevenTwo  & Master   & Principal architect / Department head  & 20 \yrs{} & 7 \yrs{}    \\
  &&& \elevenThree{} \it{(excluded)}  &\\

  \hline
  \cTwelve{} & E-mobility & Large 
  & \twelveOne  & Technical high school  & Product manager / Senior developer   & 25 \yrs{} & 5 \yrs{}         \\

  \hline
  
\end{tabular}
\end{adjustbox}

\end{table*}

\section{Message Exchange Techniques (RQ1)}
\label{sec:res1}

This section presents the message exchange techniques used in practice and their corresponding documentation techniques, which we elicited with the communication questions of our interview guide.
Hence, this section answers RQ1: \textit{\rqOne{}}

\subsection{Answer to RQ1}
\begin{figure}[ht]
    \centering
    \begin{adjustbox}{width={\columnwidth},totalheight={\textheight},keepaspectratio}%
    \includegraphics{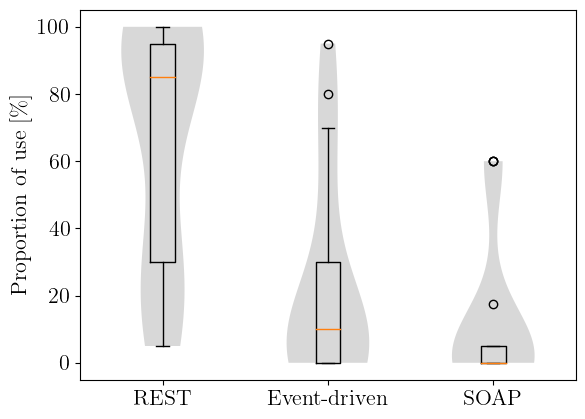}
    \end{adjustbox}
    \caption{Proportion of communication techniques of the total communication among interview participants.}
    \label{fig:comm-techniques}
\end{figure}

The two most popular message exchange techniques among the participants are Representational State Transfer (REST) APIs and event-driven communication. On average, REST APIs make up 66.8\% of the total communication \showstats{5\%}{100\%}{66.8\%}{85\%} and event-driven communication makes up 22.6\% of the total communcation \showstats{0\%}{95\%}{22.6\%}{10\%}.
Some participants %
provide and maintain Simple Object Access Protocol (SOAP) APIs \showstats{0\%}{60\%}{12.8\%}{0\% of the total communication}. However, they no longer develop new SOAP APIs but only maintain existing ones for legacy consumers and plan to discontinue them once all consumers migrated. 
Figure~\ref{fig:comm-techniques} visualizes the proportion of the three  communication techniques among the interview participants as violin plots.
All participants use OpenAPI and Swagger tools to document their REST APIs automatically. Additionally, many participants manually supplement this documentation with wiki pages or in-line source code documentation. 
The participants refrain from documenting the event-driven communication formally because it targets system-internal services with well-known maintainers.

Notably, we heard of specialized protocols such as GraphQL\footnote{\url{https://graphql.org/}} for API querying, Websockets for bidirectional communication, and Google Protocol Buffers\footnote{\url{https://protobuf.dev}} for serialization. However, only a maximum of two participants mentioned them, and hence, we did not include them in the detailed report.
In the following, we present the details of the two main communication techniques.

\subsection{Representational State Transfer (REST)}
All interview participants (17/17) provide REST APIs for their services and transfer messages serialized into JavaScript Object Notation (JSON). %

\subsubsection{REST APIs}
The participants consider REST APIs a de facto standard for service communication. They are easy to use and require little setup time for consumers, considering most developers are already familiar with REST. 
Furthermore, most REST API frameworks support authorization protocols such as OAuth 2.0 out-of-the-box. %
Hence, many participants (9/17) %
exclusively use REST for public-facing APIs to customers.
A few participants (3/17) provide client SDKs abstracting the REST calls, but this approach increases the maintenance overhead with each additionally supported development language.

REST~\citep{fielding2000rest} is the de facto standard to publicly expose request-response APIs for accessing data and computing resources of web services~\citep{KRATZKE20171,zimmermann_et_al:OASIcs:2020:11826}.
REST utilizes the hypertext transfer protocol (HTTP) to exchange messages, and consumers of REST APIs use unique resource identifiers (URIs), called \textit{endpoints}, to request domain objects called \textit{resources}. The HTTP methods indicate the request's \textit{operation}, e.g., \texttt{GET} for read, \texttt{POST} for create, \texttt{DELETE} for delete. 
Similarly, the HTTP status codes directly indicate the response type, e.g., for the \texttt{GET} request, the service might return a \texttt{200} with the requested resource or a \texttt{404} if it was not found. 
According to \cite{KARABEYAKSAKALLI2021111014}, REST APIs have the advantage of easy implementation but require well-refined request-response data structures and the availability of both the provider and consumer services at call time to process a request correctly.

\subsubsection{REST API gateways}
\label{comm:restgateway}
Many participants (9/17) %
implement dedicated API gateways that handle all incoming REST API requests. Their API gateways abstract the individual services' APIs and versioning by hiding the internal architecture and only providing a single access point for external consumers, enabling system transparency and loose coupling. Further, the API gateways centrally manage the authentication of requests and eliminate the redundancy of implementing it for each service's API individually. 

A REST API gateway implements the \textit{Facade} pattern~\citep{10.5555/186897} on component level~\citep{10.1007/978-3-319-69035-3_29}.
\cite{closer18} recommend the API gateway as an extensible and backward-compatible orchestration and coordination pattern, and 
an MSA without an API gateway is considered bad practice~\citep{8354414, 8890660}.
As a disadvantage, the single API gateway is a potential bottleneck. Load balancing techniques~\citep{closer18} and resiliency patterns~\citep{9101301}, such as Retry and Circuit Breakers, mitigate this disadvantage but increase the development and runtime complexity.

\subsubsection{OpenAPI and Swagger documentation}
\label{docs:openapi}
All participants~(17/17) %
use the OpenAPI specification\footnote{\url{https://spec.openapis.org/oas/v3.1.0}} to define and document their REST APIs formally.
The OpenAPI specification allows the clear documentation and versioning of the REST API. 
The participants share the OpenAPI specification also with external consumers, who are typically familiar with the format or even use OpenAPI themselves.
The participants use Swagger tools\footnote{\url{https://swagger.io/tools/}} to automatically generate and visualize the OpenAPI specification in the browser, including directly executable REST call examples.
Some participants~(5/17) %
use the OpenAPI and Swagger capabilities to automatically generate server code, consumer code, contract tests, and client SDKs.

OpenAPI is a vendor-neutral description format by the Linux Foundation and de-facto standard in the industry. 
\cite{8385157} analyzed $500$ REST APIs and found that almost half automatically generate the specification with Swagger.
Various practical tools\footnote{\url{https://openapi.tools}} and research approaches~\citep{10.1145/3184558.3188740, PENG20181313, 10.1007/978-3-030-50578-3_40, 9159071} utilize the OpenAPI specification format.
However, the OpenAPI specification only describes the API structure, not the API behavior, e.g., authentication and relations between message fields.

\subsubsection{Supplementary manual documentation}
Many participants~(9/17) %
supplement the OpenAPI documentation with manual documentation about the REST API behavior in written form or UML diagrams.  
The participants use wiki pages, e.g., Confluence (6/17), %
to manage the supplementary documentation, where they also link to previous API versions and the OpenAPI specification.
\citeinterview{It's not enough to just show the REST interface and the parameters, you have to know some business context around it}{\threeOne}
The supplementary documentation explains the authentication processes, message fields' semantics and relationships, and error handling and recovering options. 
For instance, the relationship between the two fields \texttt{balance} value and \texttt{tax} flag could vary. If the tax flag is set, it could mean the tax value was added to the balance. Alternatively, it could mean that the balance is deductible.
Interestingly, a few participants (4/19) manually document the REST API structure, e.g., REST endpoints and parameters, and provide example calls, e.g., in curl. 
Swagger tools could generate such documentation as OpenAPI documentation automatically, however, in a more technical format.

While OpenAPI serves as a specification for a REST API's structure and input and output formats, the API's semantics are often documented in natural language or even missing. The resulting ambiguity of semantics complicates integrating multiple services with different contexts and domain vocabulary \citep{10.1007/978-3-319-67262-5_12}.
\cite{8029798} proposed an approach to derive the semantics by semi-automatically matching the OpenAPI structure with public ontology concepts.

\subsubsection{Internal source code documentation}
\label{comm:internaldocs}
Notably, some participants (6/17) %
do not document system-internal REST APIs on the API level. They prefer reading the source code and in-line code documentation directly, %
especially if the whole code base is already loaded in the editor. They consider it faster than loading the Swagger-generated documentation and manually identifying the semantics and behavior.

This documentation strategy for system-internal functionality follows conventional development practices and is unrelated to the MSA. Hence, we refer to conventional source code documentation research, e.g., \cite{10.1007/978-3-319-19243-7_10}.

\subsection{Event-driven communication}
Many participants (13/17) %
use event-driven communication patterns, such as publish-subscribe and message queues, system-internally to send asynchronous messages via message brokers. 

\subsubsection{Asynchronous messaging and message brokers}
The participants %
use asynchronous messaging whenever real-time responses are not required and eventual consistency is acceptable, e.g., on state updates or completion notifications of long-running processes. 
\citeinterview{It's the cloud. It's async anyways, why not make it explicit?}{\fiveOne}
The participants mainly use the two message broker technologies RabbitMQ\footnote{\url{https://www.rabbitmq.com}} (6/17) and Apache Kafka\footnote{\url{https://kafka.apache.org}} (3/17). 
This layer of abstraction loosely couples the system. New services are easily added and removed as publishers and subscribers without adapting any other services. Similarly, asynchronous messaging helps to integrate services with an existing monolith by notifying them about internal events without altering the  monolith's original program flow.

Asynchronous communication via message brokers, a type of simple message-oriented middleware~\citep{8905013, 8560493},
decouples services during development, deployment, and runtime. 
RabbitMQ and Apache Kafka are two of the most popular message broker technologies~\citep{8905013}. 
Message brokers automatically distribute messages based on the messages' \textit{topics}. They automatically \textit{broadcast} \textit{published} messages to all currently \textit{subscribed} services or store them in a \textit{message queue} for immediate or later consumption.
Contrary, calling a REST API requires the explicit knowledge and availability of all directly called services. 
As a downside, introducing message brokers increases the system's complexity because they hide communication paths and dependencies between the services~\citep{KARABEYAKSAKALLI2021111014}.

\subsubsection{Limited documentation of event-driven communication}
\label{docs:event-driven}
Contrary to REST API documentation, only a few participants (3/17) %
explicitly document the event-driven communication, and only one of them uses the AsyncAPI\footnote{\url{https://www.asyncapi.com/docs/reference/specification/v2.6.0}} specification. 
We theorize most participants do not formally document their event-driven communication because it aims at system-internal communication. Hence, the audience for such documentation consists mainly of other company-internal developers with access to the source code (cf. Section~\ref{comm:internaldocs}). 
\citeinterview{Actually, it's quite easy to look into the services to get an idea how the payloads look like and so on}{\fiveTwo}
In contrast, REST APIs are open to external consumers and customers with less technical or domain-specific backgrounds. 

AsnycAPI is the de-facto industry standard for documenting message-based communication~\citep{10.1145/3365438.3410948}. Interestingly, it is not well recognized by our interview participants who instead rely on the source code.
\cite{KARABEYAKSAKALLI2021111014} stated that the advantage of the MSA is eliminated once the system dependencies cannot be handled anymore. 
Accordingly, dynamic monitoring approaches such as Helios~\citep{6062132, 10.1145/2335484.2335511} and D\textsuperscript{2}Abs~\citep{7582771, 9618848} are used to identify or recover the dependencies and potential change impacts between services.

\section{API Evolution Strategies (RQ2)}
\label{sec:res2}
This section presents the API evolution strategies that we found with the provider and consumer API evolution questions and the additional thoughts discussions of our interview guide.
It answers RQ2: \textit{\rqTwo{}}

\begin{table}[t]
  \centering
  \caption{%
   The API evolution strategies with Participant and Company counts. 
   }
  \label{tab:strategies}

\begin{adjustbox}{width={\columnwidth},totalheight={\textheight},keepaspectratio}%
\begin{tabular}{lrr}

\hline
API Evolution Strategy                     & \# P & \# C \\
\hline 
\hline

Accept necessary breaking changes          &  17 &  11 \\
\hspace{1em} Understand the reasons for breaking changes & 17  & 11   \\
\hspace{1em} Consider structural and behavioral changes  & 5  & 4   \\
\hline

Stay compatible and avoid unexpected breaking changes &  17 & 11  \\
\hspace{1em} Work around breaking changes  & 17  & 11   \\
\hspace{1em} Regression test the API       & 10  & 8   \\
\hspace{1em} Think ahead and design a dynamic API  & 6   & 6   \\
\hline

Version the API                            &  17  & 11  \\
\hspace{1em} Create a new version on breaking changes  & 17  & 11  \\
\hspace{1em} Expose multiple versions simultaneously    & 13  & 8   \\
\hline

Collaborate with other teams           &  15 &  9  \\
\hspace{1em} Actively involve consumer teams & 14  & 8 \\
\hspace{1em} Follow the API-first approach & 11  & 8   \\
\hline 

Internally, just break (and fix) it        & 11  & 10  \\
\hline

Abstract external systems' APIs        & 6   & 5   \\
\hline

\end{tabular}
\end{adjustbox}

\end{table}

\subsection{Answer to RQ2}
The interview participants apply five strategies to evolve the provided microservice APIs and one to handle the evolution of consumed APIs. Table~\ref{tab:strategies} contains the complete list formulated as comprehensive best practices that practitioners should follow when evolving microservice APIs.

First off, all participants must deal with breaking changes from adding or improving the functionality and system maintenance efforts.
All participants stay compatible with existing consumers and actively avoid introducing unexpected breaking changes. Many participants apply regression testing to detect unintentional breaking changes before release. Some participants implement dynamic APIs allowing custom queries where consumers decide the message fields in the response. 
All participants version their APIs and indicate breaking changes with increased version numbers. Many provide multiple API versions allowing consumers to migrate at their own pace.
Most participants collaborate with dependent teams by discussing the planned API changes before implementation. They focus on the API definition before implementing the underlying functionality in parallel.
Many participants agree that system-internal changes without impact on the public API do not require special handling, e.g., formal planning or versioning.
Finally, some participants promote an abstraction layer for external systems that handles authentication and message translations.
In the following, we present the details of each strategy.

\subsection{Accept necessary breaking changes}
\label{strat:acceptbc}
According to Lehman's laws of software evolution~\citep{LEHMAN1979213}, real-world software systems require maintenance and evolution to stay relevant. Consequently, all participants (17/17) must deal with breaking API changes.

\subsubsection{Understand the reasons for breaking changes}
From the interviews, we identified three main reasons for breaking changes: 
a) introducing new functionality (12/17)%
, e.g., extending existing workflows or providing more diverse workflows and APIs, 
b) improving existing functionality (9/17)%
, e.g., merging similar workflows or changing the underlying technology, and 
c) improving the API design (6/17)%
, e.g., removing outdated workflows or restructuring the exposed API.
Other reasons include bugfixing (4/17), %
introducing or changing security and authentication techniques (4/17), %
migrating to a changed external system (4/17), %
and changes to the underlying infrastructure, e.g., migrating to a new message broker or cloud provider (4/17). %
The participants introduce breaking changes only quarterly to half-yearly to provide enough lead time for affected consumers. Two exceptions are bugfixes and API migrations, which require timely breaking changes to retain a stable system again.

\cite{7884616} analyzed changes in Java library APIs and identified $28\%$ as breaking.
\cite{Brito2020} found three main motivations for breaking library API changes, which are similar to our findings: implementing new features, simplifying the API, and improving maintainability. 
\cite{6649592} stated that more than 80\% of web service API changes are refactorings, matching our findings for improving the existing functionality and API design.

\subsubsection{Consider structural and behavioral changes}
Our participants described two types of breaking changes: structural and behavioral.
All participants (17/17) considered structural changes, e.g., deletions and renamings, breaking changes.
Interestingly, only a few participants~(5/17) %
identified behavioral changes, e.g., changing a timestamp's timezone or returning unexpected values, as harmful and handled them as potential breaking changes. \citeinterview{The customer still recognizes that the values changed from last time and then triggers a support ticket to ask about it}{\threeTwo}
Others intentionally introduce behavioral changes to avoid structural changes and use organizational strategies as justification, e.g., declaring all fields optional (cf. Section \ref{strat:dynapi}).

\cite{Newman:15:MS} called the two breaking API change types structural and semantic, and the service's internal behavior influences the semantics. 
\cite{https://doi.org/10.1002/smr.328} considered behavioral changes in Java libraries breaking because changed computation results require different consumer-side handling. 
Similarly, \cite{Fokaefs2014} classified web API changes as no-effect, adaptable, and non-recoverable, representing internal, structural, and behavioral changes, respectively. They justified their terminology because structural changes are recoverable by adding wrappers around consumer services, but behavioral changes require code changes and re-deployment to handle the changed computation results.

\subsection{Stay compatible and avoid unexpected breaking changes}
\label{strat:compat}
The main strategy followed by all participants (17/17) is to avoid unexpected breaking changes and to stay backward compatible. \citeinterview{We ensure that we don't have any breaking changes, unless it's absolutely necessary}{\twoTwo} The participants do not actively notify API consumers about backward compatible changes, except for the persons requesting the new functionality.

\subsubsection{Work around breaking changes}
Whenever possible, the participants (17/17) plan and implement workarounds to avoid breaking changes. 
\citeinterview{We would discuss this, what it means, what is affected by it, and then we try to find a solution that does not change something for the [existing consumers]}{\tenOne}
Many participants (11/17) recommend extending an API by adding or duplicating endpoints, messages, and fields to ensure compatibility of the new functionality with the existing consumers. 
One participant %
warns of introducing tailored APIs for a single use case to avoid breaking changes in others. In their experience, this fragments the communication interfaces and increases the system's complexity.
Similar to the individual APIs, from a system perspective, the business workflow should not break after introducing changes. For instance, a business workflow of buying a product has multiple operations: i) adding the product to the shopping cart, ii) legally purchasing the product, and iii) paying for the product. The order and results from executing these operations should stay compatible with all consumers.

Theoretically, this strategy provides the best results for providers, who only maintain a single API version, and consumers, who do not need to change their implementation. In practice, avoiding breaking API changes is not always possible~\citep{7884616, Brito2020}.
\cite{daigneau2011service} advocated the \textit{Tolerant Reader} pattern, e.g., only accessing needed message fields, not relying on orders but identifiers, and wrapping domain-specific objects in more general structures such as lists and maps, to reduce consumers' susceptibility to breaking changes.

\subsubsection{Regression test the API}
Many participants~(10/17) recommend regression testing to detect accidentally introduced breaking changes before re-deploying the services. %
\citeinterview{You already have a set of test cases that you can run against the old interface. You will see immediately when you introduce something that breaks it}{\twoThree}
\citeinterview{Before releasing, we run each and every test we have - and this is a quite huge test suite - over the last release version again to make sure nothing did break in between, since the last release}{\fiveTwo}
Unit testing the source code detects behavioral changes, e.g., changed result values.
Contract tests detect structural and behavioral changes in the API, e.g., required parameters or unexpected response objects. 
Finally, a few participants (3/17) execute complete end-to-end tests to ensure functionality and backward compatibility for the most important workflows. %
When introducing planned breaking changes, the loose coupling of services requires the developers to adapt their contract tests to the changes manually. \citeinterview{At the point of writing, my test data has the format that you would say it will have. If you change it, then I will have to change my test data as well}{\sixOne}

Regression testing is an established practice to raise confidence that program modifications have no unexpected adverse effects~\citep{65194, 630875}. 
\cite{biswas2011regression} concluded that regression testing component-based systems helps to detect indirectly modified APIs after behavioral changes in the business logic. 
While only a few participants stated the types of tests they employed, we expect them to use the functional tests for MSA recommended by~\cite{richardson2018microservices}:
unit tests, %
integration or contract tests, %
component tests, 
and end-to-end tests. %
\cite{https://doi.org/10.1002/spe.2967} identified test case generation as an open concern in grey literature. %
\cite{10.1145/3395363.3397374} proposed an approach for regression testing structural and behavioral REST API changes by automatically generating requests and comparing the responses for multiple service and consumer version combinations. 
Frameworks, such as Spring Cloud Contract\footnote{\url{https://spring.io/projects/spring-cloud-contract}} for Java and pact-net\footnote{\url{https://github.com/pact-foundation/pact-net}} for .NET, simplify REST API testing. They generate server stubs for consumer-side testing and define a domain-specific language to write requests for server-side tests.
\cite{electronics11172671} proposed regression testing for message-driven APIs. They extracted low-level TCP and UDP package payloads and reverse-engineered the request and response messages into future regression test cases.

\subsubsection{Think ahead and design a dynamic API}
\label{strat:dynapi}
Some participants~(6/17) recommend designing dynamic APIs with the goal of a flexible API resulting in fewer breaking changes.
Dynamic APIs publish all available fields of a response object, and consumers pre-filter them as part of the request. 
The consumers then only receive their subset of fields, potentially containing null values.
The API developers must plan ahead and consider current and future use cases and their expected responses to allow such dynamic APIs.
\citeinterview{Therefore, in general, we [...] think ahead and we try to add many times also attributes in advance}{\nineOne}
With this approach, the developers design a clear, extensible, multi-purpose API for the underlying functionality instead of a specific API tailored to one use case. 
The participants recommend JSON objects compared to strings or binary because JSON allows for hierarchies, lists, and null values.
A few participants~(3/17) %
declare most fields optional to avoid future breaking changes. 
Surprisingly, only one participant mentioned GraphQL\footnote{\url{https://graphql.org/}}, a query language specialized in querying a subset of response fields. The others implemented the dynamic APIs with REST. 

\cite{10.1145/1176617.1176622} advised self-explanatory and extensible APIs. They should not overconstrain but serve multiple use cases.
Consumers adhering to the \textit{Tolerant Reader} pattern~\citep{daigneau2011service} can react to changes in list sizes, hierarchical structures, and null values of APIs gracefully and might even recover from moved fields. 
\cite{9101226} showed that GraphQL queries required less implementation time than REST, with improved results for increased query complexity. \cite{8667986} found that client-specific GraphQL queries allowed a reduction of JSON response fields by $94\%$ compared to REST. 
\cite{10.1007/978-3-030-33702-5_1} revealed exponential response times and sizes for GraphQL queries in practice and recommended throttling requests and pagination techniques. 
Similarly, \cite{10.1145/3561818} concluded that GraphQL requires more best practices and improvements in query complexity, code generation, and security. 
These findings might explain our study results, where only one participant used GraphQL while others preferred custom REST implementations.

\subsection{Version the API}
\label{strat:versioning}
All participants (17/17) apply versioning to evolve their APIs and use the version information in requests and messages to access the corresponding API version of a service. \citeinterview{We have defined that for every API that is accessible from the outside we do have versioning}{\threeOne}
Notably, the participants focused on REST APIs when discussing versioning and we identified versioning for event-driven communication as a challenge (cf. Section~\ref{chall:version-event-driven}).

\subsubsection{Create a new version on breaking changes}
All participants (17/17) increase the API version when introducing breaking changes. Non-breaking changes, such as exposing new endpoints or extending message objects, are implemented in the latest API version directly. 
A few participants~(5/17) %
mentioned semantic versioning\footnote{\url{https://semver.org}} explicitly, but only the major version number indicates breaking API changes and is relevant for consumers. Hence, the remaining participants~(12/17) %
simply use increasing integer values for API versions, e.g., $v1$, $v2$. 
Independently of the internal versioning granularity, the REST API endpoints and event-driven communication topics only contain the major version number to indicate compatibility.

Semantic versioning or integer versioning are well-known strategies for indicating breaking changes in API management~\citep{8944981, 9426799}. 
\cite{8385157} analyzed 500 REST APIs and found that $65.4\%$ exposed the major version within the request call.
Similarly, \cite{10.1007/978-3-031-34444-2_22} analyzed $7,114$ REST APIs, and the majority used static versioning in the URI or request metadata ($70.1\%$) or dynamic version discovery through a dedicated endpoint ($3.1\%$). %
\cite{8354414} identified not having API versioning as an MSA smell.

\subsubsection{Expose multiple versions simultaneously}
\label{strat:parallelvers}
Ideally, a new API version supersedes the previous version, and development teams only maintain the latest version as a single source of truth.
In reality, many participants (13/17) expose multiple API versions simultaneously to serve consumers who do not or only infrequently update their API calls.
The newest features are only available in the latest API version and all consumers requiring these features must update their calls. Other consumers are unaffected and continue using the previous API versions. 
We found two approaches for running multiple API versions in parallel: 
exposing all API versions in the same service instance (8/17), %
and deploying each service version separately (5/17). %
Some participants (8/17) consider exposing all API versions within one service easier to maintain because the underlying business logic stays consistent. \citeinterview{So just this simple mapping of DTOs. When you use the right technologies it's quite OK and not that much of effort}{\twoOne}
In contrast, deploying each service version and API separately duplicates the source code base and requires more complex message routing. Still, some participants (5/17) prefer the smaller service instances.
The number of simultaneously exposed API versions typically ranges from $2$ to $8$.
\citeinterview{We have to keep the last three versions running, not more}{\eightOne} 
\citeinterview{For core systems we have about 7 to 8 breaking versions}{\twoOne} 
Still, the participants only remove old APIs once all consumers migrated to the newer version. After all, they must support all customers independently of the request versions. This sometimes requires the participants to support old API versions indefinitely, especially for important and slowly responding customers~(cf. Section~\ref{chall:consumer-compat}).

\cite{Newman:15:MS} proposed the same two strategies we found for running multiple API versions in parallel: \textit{emulating the old interface}, i.e., exposing all API versions in the same service instance, and \textit{coexisting incompatible microservice versions}, i.e., deploying each microservice version separately. Like our participants, he recommended emulating the old interface because this approach is easier to maintain, evolve, and monitor.
\cite{8385157} reported that about two-thirds of the $500$ analyzed REST APIs supported API version selection, indicating multiple active versions.
The \textit{Parallel Change} pattern~\citep{parallelchange2014} requires both the old and new versions running for the consumers to migrate at their own pace. Providers remove the old version once the consumers finish the migration.
\cite{10.1007/978-3-662-45391-9_17} found that web APIs follow such deprecate-replace-remove cycles in practice.
\cite{10.1007/978-3-031-34444-2_22} encountered $135$ out of $7,114$ REST APIs with multiple active versions and a maximum number of $14$ coexisting versions. 
We explain the low number of $135$ compared to our qualitative result with their automated extraction approach. They extracted the version information from the OpenAPI specifications, where we expect providers to motivate consumers to use the latest version (cf. Section~\ref{chall:parallelv}).

\subsection{Collaborate with other teams}
\label{strat:collab}
Most participants (14/17) closely collaborate with teams of consumer services during the API evolution process by providing change previews, receiving early feedback, and synchronizing integration. While they are finally responsible for their APIs' evolution, they value consumers' feedback.
\citeinterview{So, if other product teams, for example, are affected by this [change], then we first have some discussion rounds about it.}{\fiveTwo}

\subsubsection{Actively involve consumer teams}
Most participants (14/17) discuss planned API changes with consumer teams and use the feedback to improve the underlying workflow and API design before release. Many participants~(11/17) %
schedule meetings for these discussions, while a few~(3/17) %
distribute the version previews for asynchronous feedback loops.
According to the participants, one or two people per dependent system are involved in the meetings, which take up to one hour. 
Once the involved teams accept the API design, they implement the services and consumers in parallel and add them to the testing environment as soon as possible for additional feedback.
Encountering problems with the agreed-on specification during the implementation phase triggers additional follow-up discussions.
Finally, the teams jointly write contract tests and plan the deployment of the individual components. \citeinterview{But we also make the meetings to describe the changes and make tests together on the QA systems and define a date where they switch over to the new interface. And we monitor if it works for them}{\threeOne}

\cite{richardson2018microservices} acknowledged that features spanning multiple services require careful coordination between development teams. 
\cite{10.1145/3447245} found that developers considered breaking changes the provider's responsibility, who felt personally obligated to help resolve them.
This close collaboration results from the loosely coupled MSA, which, by itself, does not provide any immediate feedback on the implementation's and integration's correctness~\citep{10.1145/1526709.1526832}.

\subsubsection{Follow the API-first approach}
\label{strat:apifirst}

To simplify the collaboration efforts, many participants (11/17) discuss and agree on the API definition with consumers before starting the implementation.
This API-first approach continuously improves the API design based on consumer feedback without having to implement the actual logic behind the interfaces. \citeinterview{It's an iterative process. There is no shame in having a final-v2}{\sixOne}
The main goal is to create a well-defined API, not a functional prototype.
According to the participants, the API-first approach improves the overall design by focusing on readable, self-documenting, and reusable APIs. 
\citeinterview{Both the [internal developers] and the customer using the public API have a nice experience and get all the same information}{\sevenOne}
Furthermore, changing a preliminary API definition requires less effort than changing a partially implemented system. 
The participants %
use the OpenAPI specification (cf. Section~\ref{docs:openapi}) to document and distribute the REST API definition when following the API-first approach. 
Some participants~(5/17) %
further use the OpenAPI specification to automatically generate server code, consumer code, contract tests, and client SDKs.

\cite{KopeckyFremantleBoakes_2014_90_97} described the API-first approach as first building the functionality as API and only then creating clients for that API. 
Hence, developers design APIs to provide their business functionality to the outside, not to support specific use cases~\citep{10.1145/3308560.3320089}. 
\cite{10.1007/978-3-030-97652-1_10} concluded that the API-first approach creates clear and well-defined APIs exposing business capabilities, reducing the domain coupling with consumers, and allowing parallel development.
\cite{10.1007/978-3-642-39200-9_4} proposed an approach to generate the API design from user interface mockups as a starting point for the development process.
Vice versa, \cite{10.1145/3543895.3543939} interviewed four developers who motivated an applicability study to automatically generate user interfaces from API definitions.

\subsection{Internally, just break (and fix) it}
\label{strat:internal-break}
Many participants (11/17) agree that internal breaking changes are easier to implement and integrate, and, hence, occur more frequently. \textit{Internally} refers to the accessibility scope of the breaking API, i.e., the affected consumers are well-known or their source code is directly accessible. 
Many developers (9/17) %
introducing breaking changes also change all the consumers and the test suites. 
Some developers (6/17) %
are in close contact with the colleagues maintaining the consumers, or directly create pull requests for the consumers' source code.
Accordingly, a few participants~(4/17) %
explicitly stated they do not version internal APIs, but update, test, and redeploy them directly. 

This evolution strategy for internal APIs is unrelated to the MSA. Hence, we refer to conventional source code evolution research, e.g., \cite{Brito2020}.

\subsection{Abstract external systems' APIs}
\label{strat:abstraction}
Finally, some participants (6/17) use dedicated \textit{integration services} to abstract communication with external systems. These services handle the authentication with the external systems and translate request and response field names to the internal domain names. The internal services then do not know about the external systems they communicate with.
This allows the integration services to partially handle breaking changes in external systems, e.g., changed authentication, moved or renamed fields, and some semantic changes, and convert them back to the expected values. Hence, they minimize error propagation, and an external API change might not affect other internal services.

The integration service is the consumer-side counterpart of the API gateway (cf. Section~\ref{comm:restgateway}).
This abstraction layer follows the \textit{Proxy} and \textit{Facade} patterns~\citep{10.5555/186897} on the component level, e.g., implementing access functionality and simplifying the external interfaces.
\cite{ESPINHA201527} conducted six interviews where the developers advised to contain external web API changes to a small set of files, and
\cite{Fokaefs2014} considered structural changes recoverable by adding a wrapper to the original consumer service. 
Similarly, \cite{Wu2016} recommended encapsulating external libraries to reduce the potential change impact.

\section{API Evolution Challenges (RQ3)}
\label{sec:res3}
\label{sec:challenges}
This section presents the API evolution challenges that the participants encountered. We elicited them with the provider and consumer API evolution categories and the additional thoughts question of our interview guide. This section answers RQ3: \textit{\rqThree{}}

\begin{table}[t]
  \centering
  \caption{%
   The API evolution challenges with Participant and Company counts. 
   }
  \label{tab:challenges}

\begin{adjustbox}{width={\columnwidth},totalheight={\textheight},keepaspectratio}%
\begin{tabular}{lrr}

\hline
API Evolution Challenge                     & \# P & \# C \\
\hline 
\hline

Manual change impact analysis is error-prone         & 14  & 11  \\
\hspace{1em} Code changes affect the API unexpectedly   & 9  & 7   \\
\hspace{1em} Understanding consumed APIs' changes is effort    & 9  & 7   \\
\hline

Consumers rely on API compatibility & 12  & 7  \\
\hline

Communication with other teams lacks clarity                  &  9 & 7 \\
\hspace{1em} Consumers might be unknown   & 7  & 5  \\
\hspace{1em} Informal communication channels   & 17 & 11   \\
\hspace{1em} Communication suffers from hierarchy    & 6  & 4   \\
\hline

API maintainability and usability degrade over time        & 14  &  9  \\
\hspace{1em} Outdated API versions add maintenance overhead & 10  & 8 \\
\hspace{1em} Backward compatibility increases technical debt  & 9  & 6 \\
\hline 

Governmental services are uncooperative      & 6 &  4 \\
\hline

Event-driven communication evolution is disregarded        & 7  & 4  \\
\hline

\end{tabular}
\end{adjustbox}

\end{table}

\subsection{Answer to RQ3}
We identified six challenges in the API evolution process, out of which three result in degrading API maintainability and usability. Table~\ref{tab:challenges} contains the complete list formulated as comprehensive pitfalls for practitioners.

First, most participants encountered problems understanding the impact of source code changes on their APIs and the impact of external API changes on their services.
Second, consumers of many participants fully rely on API compatibility and refrain from migrating to a new version.
Third, many participants considered communicating with other teams challenging, especially for company-external teams. They followed no general communication strategy and suffered from hierarchical communication.
As a result of these challenges, the API cannot evolve sustainably, and most participants report degrading API design and increasing technical debt.
In contrast, governmental services choose to evolve APIs regardless of consumer concerns, which poses a challenge for some participants.
Finally, we noticed that participants hesitated to discuss event-driven communication, and some deemed evolving event-driven communication challenging.
In the following, we present the details of each challenge.

\subsection{Manual change impact analysis is error-prone}
\label{chall:manualCIA}
Most participants (14/17) find assessing source code and API change impact challenging. Based on the service boundaries, we split this challenge into two: the impact of source code changes on the provided APIs and the impact of changes in consumed APIs on the source code.

\subsubsection{Code changes affect the API unexpectedly}
Many participants (9/17) state that development teams must manually assess the impact of source code changes on the API.
They experienced that the developers sometimes overlooked that they introduced breaking changes to the API and published them without versioning or notifications. 
\citeinterview{From time to time we face problems, but mainly because some team has overlooked that it has been doing a breaking change}{\twoOne}
Consequently, one participant %
uses git diff to extract the changes between two external OpenAPI specification versions manually to identify overlooked structural breaking changes.
However, a few participants (4/17) %
consider behavioral changes especially challenging because
developers lack the tools to identify their impact automatically. Static analysis tools have problems detecting behavioral changes, e.g., changes in the return values of methods, 
and automated tests cannot cover all execution paths. \citeinterview{When you go into this [...] topic, it's not so easy to test all constellations}{\tenOne} 

Static analysis tools, e.g., openapi-diff\footnote{\url{https://github.com/OpenAPITools/openapi-diff}}, extract structural changes between two OpenAPI specification versions. 
However, \cite{7886973} conducted $35$ interviews with software developers and reported that integration challenges mainly related to semantic and behavioral changes introducing unpredicted side effects.
\cite{10.1007/978-3-030-04771-9_45} proposed model-driven microservice development. They assessed the impact of model changes by assembling the system to execute integration tests and marked the conflicting microservices.
\cite{MA2019724} automatically prioritized contract and unit tests based on the service dependencies to identify and prevent unexpected breaking changes in the MSA faster. 
\cite{8918947} introduced a control and data flow analysis technique to extract the semantic change impact from code without relying on test execution.
\cite{8328911} applied web service slicing by identifying changed WSDL operations based on the source code's behavioral changes and used the slice for regression test selection.

\subsubsection{Understanding consumed APIs' changes is effort}
\label{chall:consumerAPIassessment}
Many participants (9/17) encountered problems with analyzing external API changes.
We found that the participants do not follow a generalizable strategy when filtering external change notifications for relevancy or assessing the change impacts, except that they do it manually. 
Sometimes (6/17), the development teams are the ones who assess the impact of external changes. The developers identify the dependencies and relevant service changes by reading the change notifications, external documentation, and own source code. In this case, understanding strongly depends on the notification and documentation quality.
\citeinterview{Because if it just says: API extension, there's a new field in there, you think, yeah, for what?}{\twelveOne} 
If the notifications contain the thoughts and reasons for the API evolution it is easier to identify, understand, and integrate relevant changes. 
\citeinterview{The only challenge then is really to find any edge cases that are not described in the documentation, and which then will cause errors in our system.}{\elevenOne}
A few participants (4/17) noted that provider teams pre-filter the breaking change notifications for the consumer teams. While this approach reduces unnecessary communication, misjudgments result in system failures. \citeinterview{They just missed out on one change, because they didn't know that it was important for us.}{\threeOne}
A few participants (2/17) rely on dedicated roles, e.g., product owners or architects, who know the service dependencies and actively inform the teams of external changes.

\cite{7884616} found that breaking Java API changes only impacted less than $3\%$ of their consumers. \cite{10.1145/3447245} discovered that most participants felt overwhelmed by the number of change notifications and considered integrating them risky. 
Many approaches build service dependency graphs (cf. Section~\ref{chall:consumer-knowledge})
which help to visualize the services' dependencies and narrow down potential change impacts. 
As a limitation, the approaches require access to the source code or service cluster at runtime which is not available for external consumers. Further, they cannot identify the actual impact of changes but only the potentially affected services and methods.

\subsection{Consumers rely on API compatibility} 
\label{chall:consumer-compat}
Many participants (12/17) report the challenge of convincing consumers to update their API calls to the new version after introducing breaking changes. The participants prefer removing outdated API versions and only focusing on the latest. Still, consumers rely on previous API versions even after receiving requests to update their calls within some timespan and, hence, hinder the clean-up process. 
\citeinterview{And then, if they changed, we can remove the old version finally. But that's always a bit more work because you have to keep the old version compatible}{\threeOne}
We found two main reasons for this reluctance: consumer teams 
do not have enough resources to update the API calls in the near future~(9/17), %
or do not prioritize changes to already working functionality~(7/17). %
\citeinterview{Most of the customers don't touch the code anymore for one year or 1.5 years if it works}{\fiveTwo}
Some participants~(5/17) %
explicitly stated they follow this \textit{never change a running system} strategy themselves and only migrate API calls if they require the new functionality.
\citeinterview{What for, I don’t need anything from 2.0 to 4.0, my world is running}{\twelveOne} %
While this strategy reduces the development effort from a consumer perspective, the same teams suffer from this slow and rigid migration strategy for their provided APIs.
Some participants~(5/17) %
force consumers to migrate their calls by turning off the outdated API version with a fixed, non-negotiable deadline, but this measure is not feasible for business-critical APIs. \citeinterview{Yeah, we're earning money with them so you cannot just say: sorry you cannot use it anymore}{\twoThree}

In the early years, web services introduced breaking API changes without versioning or with short deprecation periods, e.g., of three months~\citep{6649592, Fokaefs2014, 10.1007/978-3-662-45391-9_17, ESPINHA201527}. 
\cite{ESPINHA201527} found that developers preferred longer deprecation periods after conducting six interviews.
\cite{8385157} reported that two-thirds of 500 analyzed REST APIs supported version selection, indicating that breaking changes in newer versions did not immediately affect old consumers.
\cite{Hora2018} studied the impact of library API changes. While more than half %
of the investigated systems were potentially affected by changes, the majority did not react and continued using the previous version. 
To prevent this behavior, \cite{DETOLEDO2021110968} suggested a clear period of support that should not be extended.

\subsection{Communication with other teams lacks clarity}
\label{chall:commToConsumers}
Many participants (9/17) encountered problems in communicating changes with other teams. Developers do not know whom to inform, forget to notify the consumer teams, or convey the change information incorrectly.
\citeinterview{If something goes wrong, it's communication}{\threeThree}

\subsubsection{Consumers might be unknown}
\label{chall:consumer-knowledge}
The teams introducing breaking changes should know which consumers are affected and how to contact the corresponding teams, but some participants (7/17) %
miss appropriate documentation of consuming services and teams. 
Consumers not informed about the breaking changes exhibit unexpected behavior or failures after the update and require manual investigation of the problem.
\citeinterview{We got a 503 - Service Unavailable and so I called the product and asked: what's the problem here? And then they told me: Oh, right, we changed the API for that}{\threeTwo}
During our study, we could not find a generalizable consumer documentation strategy to recommend. %
Some participants~(8/17) %
rely on their teams' or managers' implicit knowledge. They discuss future API changes and potentially affected consumers with these colleagues, team leads, or architects. 
\citeinterview{He just remembers most of the time. Or maybe he has some documentation on his end. I'm not really $100\%$ sure}{\threeThree}
As a downside of this implicit knowledge, the information is lost if the individual leaves the company. 
Some participants~(6/17) %
log REST API calls from consumers with tracing tools, e.g., Dynatrace, Grafana, or custom implementations. This allows them to look up all consumer IPs or hostnames for each API endpoint and provides information about its use.
A few participants~(4/17) %
use the credentials for authenticating the calls to their services to maintain a list of actively used APIs and corresponding consumers. 
A few participants~(3/17) %
even maintain manual documentation about the consumers and contact partners for each microservice. 
\citeinterview{We document it in lists and I think this is not really ideal. So, which API is used by which. This is especially error-prone if we have to change something}{\eightOne}

\cite{DETOLEDO2021110968} recommended tracking internal and external users to directly request migrations.
Consequently, related works build service dependency graphs (SDGs) by analyzing the source code or runtime behavior of services~\citep{app11177856}.
\cite{7081882} proposed static analysis to construct a cross-service call graph by analyzing web service calls for SOAP services.
Similarly, \cite{MA2019724} visualized and analyzed SDGs by statically extracting REST API calls from the source code. 
Dynamic analysis approaches~\citep{10.1007/978-3-030-23502-4_20, 10.1145/3368089.3417066} trace the service calls at runtime to generate the SDG and analyze behavioral changes and performance issues over time. 
\cite{7582771} proposed identifying method dependencies based on the execution order of event-driven messages at runtime.
Similarly, Helios~\citep{6062132, 10.1145/2335484.2335511} and D\textsuperscript{2}Abs~\citep{7582771, 9618848} identified dependencies and potential change impacts of services at runtime by analyzing event-driven message handling and related method invocations.
As a limitation, these approaches require access to the source code or runtime environment of the services, which is not available for external partners.

\subsubsection{Informal communication channels}
We could not identify a generalizable strategy to inform consumer teams about API changes. The participants (17/17) either follow their own ad hoc strategy or accept the overhead for manual communication. \citeinterview{Sadly, there is no company standard for [communicating] versioning interfaces. It's up to the products to handle this}{\threeOne} 
Internally, the main means for written communication are e-mails (9/17), %
followed by announcement channels (6/17) %
and instant messages (4/17)%
, e.g., via Slack, Microsoft Teams, Mattermost. 
\citeinterview{And we do have people that actively have to look at these propagated changes. Are they relevant for the services I'm responsible for?}{\sixOne}
Alternatively, API changes are verbally announced in formal meetings (5/17)%
, e.g., coordination or sprint review meetings, 
and informal meetings (5/17)%
, e.g., coffee talks.
Externally, some participants~(6/17) %
use e-mails to communicate with partners and customers. 
Some participants~(6/17) %
also mentioned dedicated roles responsible for communicating and managing the API changes. 
This role could belong to the product owner, %
a dedicated coordinator position, %
or even a dedicated team centrally managing the company's API integration. %
A few participants (4/17) %
notify breaking and non-breaking API changes via release notes but simultaneously consider the natural language description too verbose for a technical assessment. 
\citeinterview{I'm pretty sure that no customer is really looking at that}{\fiveOne}
A few participants~(3/17) %
have to actively check for breaking changes, especially for larger API providers like Amazon Web Services.

\cite{ESPINHA201527} identified e-mails as the main communication channel but found developers considered them unreliable. Additionally, large providers, e.g., Google and Twitter, sent upcoming changes via e-mail lists of registered accounts.
Similarly, \cite{10.1145/3447245} reported that developers communicated pre-release announcements via e-mail and Twitter.
\cite{7196531} identified four communication channels: the API homepage, the API response, e.g., deprecation information in the header, customized e-mails, and newsfeeds.
However, \cite{9240687} found only three out of $1,368$ analyzed REST APIs proactively informed callers about deprecation in the response objects, where developers would directly see it during development or in log files.

\subsubsection{Communication suffers from hierarchy}
Some participants (6/17) %
suffer from a high level of organizational abstraction, hindering effective communication.
Communication with unfamiliar teams or external partners involves multiple developers, team leads, and company representatives, possibly altering the information with every pass down the chain. 
\citeinterview{The problems arise when too many third parties are involved because it's like the telephone game. You pretty much get completely different results at the end}{\threeThree}
Also, the involved people might forget the details of the API changes or mistakenly consider them unimportant. 
\citeinterview{Finally, we need to ask all the time for changes or if they are changed and then there is a quite big delay until we can continue.}{\nineOne}

\cite{doi:10.1080/08874417.2018.1520056} conducted interviews with $19$ software architects and reported API change communication and coordination with related services as challenging.
\cite{7886973} found that collaboration and software quality depended on the social boundaries of developing companies.
We refer to Conway’s law~\citep{conway1968committees} stating that a system's structure follows the companies' communication structure.

\subsection{API maintainability and usability degrade over time}
\label{chall:degradingdesign}
As a result of the previous challenges, many participants (14/17) experienced degrading API and source code quality. Consumers relying on a specific version and uncertainty about introducing breaking changes force providers to maintain the outdated versions and increase the technical debt with no clear resolution strategy.

\subsubsection{Outdated API versions add maintenance overhead} 
\label{chall:parallelv}
Many participants (10/17) mentioned the overhead of maintaining old API versions to ensure backward compatibility with existing consumers. 
\citeinterview{That's a huge pain for us}{\fiveTwo}
Some participants (5/17) considered the additional routing and handling logic based on the respective message version as an overhead. This logic converts the REST API requests and event-driven messages to the newest format or forwards them to the corresponding workflow version. 
When running coexisting incompatible microservice versions (cf. Section \ref{strat:parallelvers}), the system requires a dedicated routing layer because the incoming requests and messages are processed by individual runtime components.
The additional routing and handling logic increases the source code size and complexity (4/17) with the backward-compatible business logic and workflows, additional tests to verify each supported version and regression test any changes, and even backported features, further complicating outdated workflows instead of removing them. 
When running coexisting incompatible microservice versions, developers must maintain multiple code bases, one for each supported version, and synchronize them accordingly.

Some participants (5/17) feel the overhead for ensuring backward compatibility interferes with developing new features. 
\citeinterview{It holds you back if you want to change some other implementation, if you want to optimize something, or implement some new feature that doesn't work with an old way of transferring data or something like that. And it also slows you down or holds you back from developing any new features}{\twoTwo}

\cite{10.1145/3447245} called the overhead to maintain obsolete code and create workarounds for compatibility \textit{opportunity cost}. This opportunity cost transforms into consumers' migration cost once the providers decide to break and clean up the interface.
\cite{ESPINHA201527} recommended providers deprecate and remove the outdated APIs at some point to avoid increasing opportunity costs.
Similarly, \cite{10.1145/3361149.3361164} proposed three deprecation patterns: eternal lifetime guarantee, limited lifetime guarantee, and aggressive obsolescence. The first pattern provides unlimited API support, the second provides a clear deadline as part of the API version release, and the third removes an outdated version with prior notice of a deadline, which is not necessarily known during release.
The three patterns balance the forces of opportunity cost and consumer efforts.

\subsubsection{Backward compatibility increases technical debt}
Many participants~(9/17) experienced that avoiding breaking changes and favoring extensions for backward compatibility
degrade the initial API design over time. 
\citeinterview{That means we need to carry all the technical debt in our SDKs and our public APIs}{\fiveOne}
Eventually, the evolved API contains multiple workflows for the same functionality, outdated fields filled by old consumers but ignored when received, optional fields only processed by some consumers, and multiple equivalent endpoints fixing typos or supporting different languages.
\citeinterview{I mean, there's a developer perspective. You want to get rid of old, not really good working stuff, but in reality you just can't}{\fiveTwo}
The technical debt increases implementation complexity for new functionality and regression testing efforts for identifying unexpected side effects. It also increases the time for new developers to understand the system.
API usability degrades as consumers try to understand the differences between duplicated workflows, requests, and fields. 
Naturally, they expect differences and hesitate to decide on one solution by themselves.
A few participants (3/17) created a new streamlined version once their APIs became too convoluted and confusing and tried to convince their consumers to move to this cleaned-up version. In the worst case, this improvement step creates yet another version to maintain.
\citeinterview{Of course, you can also deprecate it. The question is if the other colleagues will also take it seriously}{\elevenOne}

\cite{DETOLEDO2021110968} identified poor REST API design as technical debt. It results in API instability, regular breaking changes, and increased difficulty in maintaining backward compatibility with newer versions.
\cite{10.1145/3447245} identified technical debt as a major driver for breaking changes from developer interviews. 
At some point, developers had to break the interface to introduce a clean version. 
Research on API maintainability and usability recommended following API standards, providing clear deprecation messages, and providing up-to-date documentation and usage examples~\citep{10.1145/3470133}.

\subsection{Governmental service providers are uncooperative}
\label{chall:governmental}
Some participants~(6/17) encountered problems with governmental services. The participants discussed multiple ministries of governments in multiple European countries.
Some~(5/17) criticize that governments do not provide a direct line of communication and contact partners are hardly available. They introduce breaking changes on short notice or do not notify consumers in advance at all.
\citeinterview{Sometimes they don't do it, they just change their service. [I found out] when the application crashed}{\oneOne}
Some participants~(5/17) experienced an unwillingness to cooperate. Governmental services shut down with a fixed date, and consumer requests are disregarded. They regularly change agreed-upon API specifications during development, and errors in the API are not investigated until consumers send an example call proving their claim. \citeinterview{You have to prove to them that they are wrong because they always say that you are doing something wrong}{\threeOne}

We explain this behavior with governments providing their services as a courtesy instead of a paid product with an underlying service agreement.  
\citeinterview{If the ministry offers an API where you can upload your tax data, you don't pay for it. They offer it}{\twelveOne}
Hence, they introduce breaking changes %
with aggressive obsolescence~\citep{10.1145/3361149.3361164} prioritizing the provider's maintenance costs over the consumers' costs.
Still, this freedom allows governments to avoid most of the challenges we identified.

\subsection{Event-driven communication evolution is disregarded}
\label{chall:version-event-driven}
Finally, we discovered that participants refrained from discussing the evolution of event-driven communication. 
Some participants~(7/19) consider it challenging to version the event-driven communication via message-oriented middleware, e.g., message queues and publish-subscribe.
The protocols do not support versioning natively, and the lightweight frameworks do not implement versioning out of the box.
Creating new topics for each version or utilizing message fields to store the version tags requires more manual intervention than versioning of REST APIs, where the frameworks automatically handle the version information in the URI or message header.
The asynchronous nature of event-driven communication requires consumers to accept old message versions even after all producers migrated, because old messages might still wait in the queue. 
Consequently, the participants either migrate all producers and consumers simultaneously and accept potential message loss~(3/17)%
, or start implementing their own version negotiation protocol
once the message volume becomes too large or external components are involved~(5/17). %
\citeinterview{So, we basically have a small protocol for this version negotiation to ensure that the systems can talk to each other}{\twoTwo}
One participant mentioned Apache Avro\footnote{\url{https://avro.apache.org/docs/1.11.1/}} to help with message serialization and versioning. %
 
\cite{DETOLEDO2021110968} discovered developers preferred complex REST API calls over event-driven messaging and classified it as an inadequate use of APIs. This indicates developers are less confident with event-driven communication.
\cite{doi:10.1080/08874417.2018.1520056} discovered that very few practitioners had experience with event-based architectures after conducting $19$ interviews.
\cite{9426799} used Apache Thrift\footnote{\url{https://thrift.apache.org}} and Apache Avro to define a custom description language for REST API message versioning and translation. This approach targets the message formats and, hence, could be adapted for messages of event-based communication in the future.

\section{Discussion} 
\label{sec:discussion}

In this section, we discuss our findings, put the strategies and challenges into relation, 
and formulate open research directions. Finally, we discuss the threats to the validity of our study.

\begin{figure}[ht]
    \centering
    \begin{adjustbox}{width={0.9\columnwidth},totalheight={\textheight},keepaspectratio}%
    \includegraphics{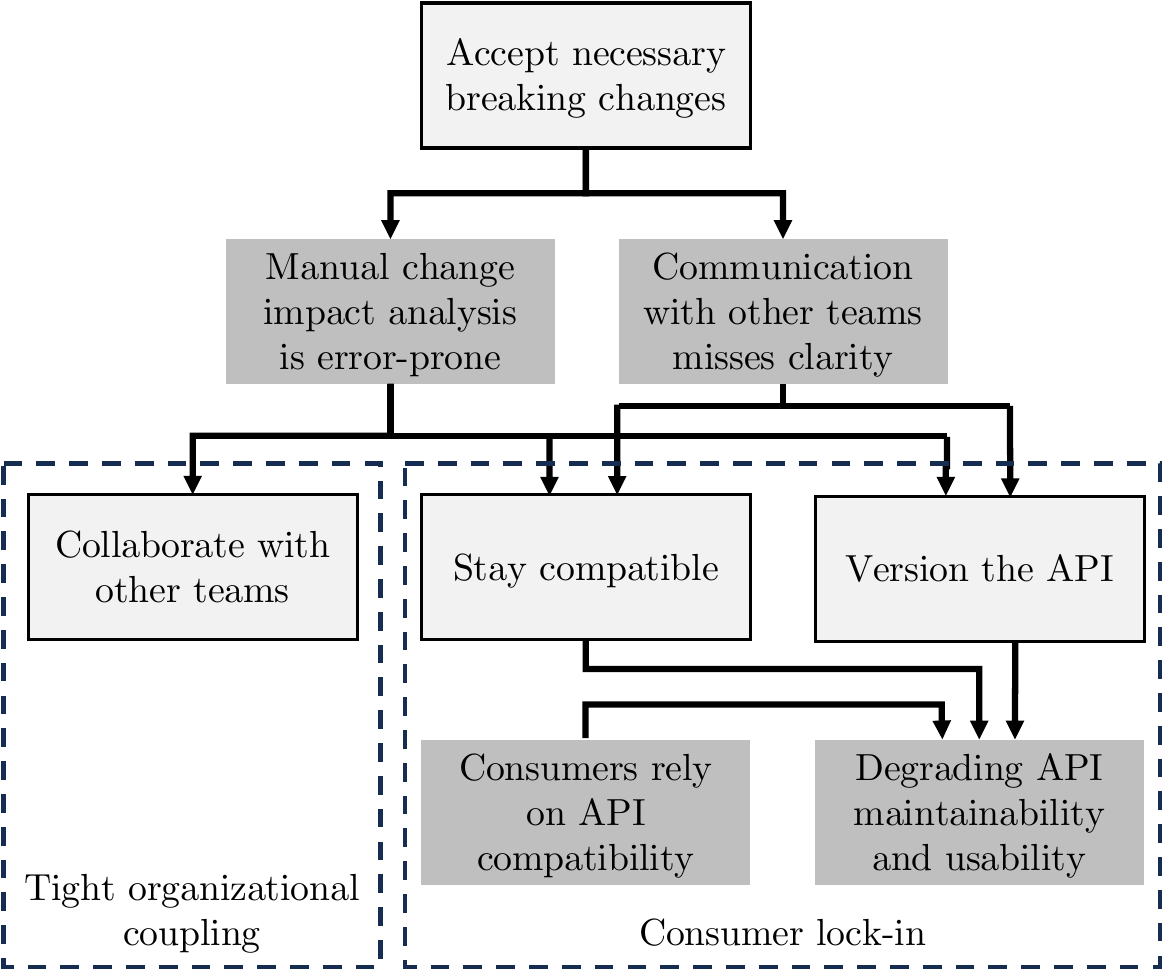}
    \end{adjustbox}
    \caption{The relationships for a subset of strategies (light grey) and challenges (dark grey) resulting in two problems (dashed).}
    \label{fig:results:relations}
\end{figure}

\subsection{Tight organizational coupling and consumer lock-in}
We identified relationships in our findings, where some strategies mitigate challenges but simultaneously raise further ones. Eventually, they result in tight organizational coupling and consumer lock-in. We visualize the relations in Figure~\ref{fig:results:relations} and describe them in the following.

While developers must accept and deal with breaking API changes during the software evolution process (cf. Section~\ref{strat:acceptbc}), they encounter challenges with understanding the impact of changes (cf. Section~\ref{chall:manualCIA}) and communicating with other teams (cf. Section~\ref{chall:commToConsumers}).
While tools like openapi-diff extract structural changes between two API versions, providers miss tools for automatically extracting API changes from changes in their implementation, especially behavioral changes, forcing them to assess the change impact manually. 
\citeinterview{They [providers] also make changes to the interface, breaking changes, without even them knowing, so they just made mistakes.
We have to tell them, hey, do you know that you changed your interface?}{\threeOne}
Further, the providers cannot fully anticipate the potential impact on consumers due to missing consumer documentation or accessibility.
The change notifications might not reach all consumers or are forgotten, e.g., when contained in a larger mail or verbally mentioned during a meeting.
If the consumers do receive a list of all the breaking API changes, they must review them and, again, assess the impact on their own system manually. 
These challenges complicate the truly independent API evolution in loosely coupled systems and organizations.
Many participants~(10/17) %
reported that their system or a consuming system broke before because the other end did not correctly assess or notify the breaking API changes.
\citeinterview{Fortunately that is quite rare, but maybe it happens once a year [in production]}{\sixOne}
\citeinterview{I would say maybe once a year, maybe twice. [...] Obviously, more often in the test system}{\threeTwo}
End users encountering the failure typically report it to the developers of the system they are interacting with and hold them accountable, independently of the root cause.

As coping strategies, providers try to stay backward compatible (cf. Section~\ref{strat:compat}), introduce breaking changes as new API versions (cf. Section~\ref{strat:versioning}), and closely collaborate with other teams (cf. Section~\ref{strat:collab}).
The close collaboration works well for cooperative teams and system-internal event-driven communication, and helps to resolve failures during integration and in production more quickly. As a downside, it shifts the overhead towards organizational communication and meetings and creates implicit knowledge distributed between team members.
\citeinterview{You see, the whole topic is really organizational-heavy, organizational and planning-heavy. [...] The technical part is then just doing it}{\twelveOne}
We call this problem \textit{tight organizational coupling}.

For external and uncooperative consumer teams, providers try to stay backward compatible and introduce breaking changes as a new API version. %
New versions increase the source code and API size and complexity by continuously adding functionality or workflows to the initial design.
A dynamic API design improves structural backward compatibility but often shifts the problem, i.e., creates semantic breaking changes instead.
\citeinterview{And these are often the problems that are only noticed once it [the system] is not working anymore. And you don't know why}{\twelveOne}
The increasing source code complexity requires extensive regression testing to ensure compatibility with old API versions. \citeinterview{Before releasing, we run each and every test we have - and this is a quite huge test suite - over the last release version again to make sure nothing did break in between, since the last release}{\fiveTwo}
These old API versions could be shut down once all consumers migrated their calls.
\citeinterview{So you have to reach out to everyone who's using your API and hope for the best. Hope that they will update}{\threeOne}
Still, many consumers rely on the API version they integrated with and do not migrate to the new version (cf. Section~\ref{chall:consumer-compat}). 
This migration reluctance forces providers to maintain an increasingly large number of API versions with every breaking change.
\citeinterview{We want to get rid of some things in our API that we already removed for newer versions or refactored for newer versions, but we just can't because this customer is still using it}{\fiveTwo}
Hence, the API cannot evolve sustainably and continuously degrades the initial API design, maintainability, and usability (cf. Section~\ref{chall:degradingdesign}).
We call this problem \textit{consumer lock-in} because the consumers force their providers to continue supporting all outdated API versions in use. 
This complicates the development of new features which would break existing calls or workflows, and increases the implementation and maintenance overhead and the technical debt with each additional API version.
\citeinterview{The cost of these workarounds that we do, I don't know. I don't dare to estimate that.}{\twoOne}

We observed that participants avoid the consumer lock-in internally, where they regularly break the API and migrate the calls themselves (cf. Section~\ref{strat:internal-break}).
Governmental services avoid tight organizational coupling and consumer lock-in by regularly introducing breaking changes (cf. Section~\ref{chall:governmental}). This gives full freedom to the provider but dissatisfies the consumers and is hence unfeasible in the context of business relationships.

\subsection{Open research directions}
We propose two open research directions aiming at mitigating both problems, tight organizational coupling and consumer lock-in.
Considering the two main causes, we propose automating the change impact analysis to improve change notification accuracy and trustworthiness and researching effective ways to communicate changes to other teams to improve notification reliability and clarity.

\subsubsection{Automating change impact analysis in MSA}
The manual change impact analysis challenge (cf. Section~\ref{chall:manualCIA}) leads to tight organizational coupling and hesitant consumer migrations resulting in consumer lock-in.
Multiple approaches constructed SDGs statically~\citep{7081882, MA2019724} or dynamically~\citep{10.1007/978-3-030-23502-4_20, 10.1145/3368089.3417066} to perform change impact analysis on the service level. As a limitation, they do not consider individual API calls or behavioral breaking changes.
Other approaches trace method invocations at runtime~\citep{7582771, 10.1145/2335484.2335511, 9618848} to construct more detailed call graphs.
Though, they require access to the runtime environment of the services, which is not available for external consumers and again would require tight organizational coupling.
A different approach proposed by~\cite{8328911} identifies changed WSDL operations based on the source code's structural and behavioral changes. 

Based on the approach by~\cite{8328911}, we motivate researchers to split the change impact analysis based on the service boundaries: first, analyzing the impact of the provider's source code changes on the provider's APIs, and second, the impact of the provider's API changes on the consumers' source code. 
This enables providers to publish a complete list of API changes and allows consumers to migrate on their own terms, reducing the tight organizational coupling. 
Further, an accurate change impact analysis mitigates the risk of unexpected changes breaking the system during migration and therefore increases consumers' trust and reduces their hesitation to migrate, which currently results in consumer lock-in.

\subsubsection{Providing effective change communication for teams}
The communication challenge (cf. Section~\ref{chall:commToConsumers}) leads to the backward compatibility requirement and consumer lock-in.
Multiple studies~\citep{ESPINHA201527, 7196531, 10.1145/3447245} identified e-mails and online platforms, e.g., Twitter and homepages, as the main communication channels for change notifications. 
\cite{10.1145/3447245} found that most developers felt overwhelmed by the number of change notifications and rather participated in planned migrations, where providers felt personally obligated to help resolve breaking changes.
\cite{Hora2018} found that deprecated library APIs producing warning messages during development caused $50\%$ more reactions than deprecated REST APIs.

Hence, we motivate the research of effective and efficient communication approaches to communicate API changes in MSA.
\citeinterview{Yeah, for example, a system where you register your APIs and where you can publish updates. For example, where you can say, hey we are removing this field, and it automatically notifies everyone that needs these APIs}{\threeThree}
Addressing the communication challenge alleviates consumer lock-in by reliably notifying affected consumers with customized change logs instead of flooding them with a generic list of changes.

\subsection{Threats to validity}
In this section, we describe the threats to the validity of our study and explain our mitigation strategies.

\subsubsection{External validity}
Our study results may not be generalizable to other teams and organizations. To mitigate this threat, we sampled practitioners from $11$ companies with various industry fields and sizes, and whereof $8$ are international companies. %
Similarly, our $17$ interview participants have diverse educational backgrounds, years of experience, and technical roles. 
We achieved this by contacting colleagues with diverse technical roles and backgrounds in multiple industry fields during the snowball sampling process.
We further mitigated the study's threat to external validity by conducting interviews until our results became stable throughout the various companies and interview partners, indicating theoretical saturation~\citep{10.1371/journal.pone.0181689}. 
We did not report on strategies and challenges mentioned by less than five participants or three companies, i.e., less than a quarter each. One challenge missed this threshold by one participant: low-quality documentation of external APIs (4/17).
Finally, we grounded our findings by connecting them to previous and related works, thereby supporting and strengthening the results.

\subsubsection{Internal validity}
Our study design may have generated incorrect results, or we may have misinterpreted participants' answers. 
We mitigated this threat by 
following guidelines for qualitative studies~\citep{10.5555/1324786, doi:https://doi.org/10.1002/9781119171386.ch22}, e.g.,
asking open-ended non-judgemental questions during the interviews, encouraging participants to speak freely, and improving the interview guide after gaining insights from previous interviews.
The second and third authors independently analyzed two random interviews and we discussed the identified codes and categories, further refining them until we reached a coder agreement. 
Finally, we shared the study results with all $17$ participants for feedback and validation. We received $13$ responses, whereof two participants had minor remarks which we incorporated, and the others fully agreed with our interpretations. 
In this paper, we only reported results mentioned by at least five interview partners from at least three companies to mitigate observer bias.

\section{Related work}
\label{sec:relwork}

In the following, we report on existing literature with a focus on studies that investigated API evolution strategies and challenges. Note, we present and discuss further literature related to the individual communication techniques, evolution strategies, and evolution challenges that we identified in our study in the corresponding sections.

API evolution is extensively studied. Though, \cite{10.1145/3470133} found that $63.9\%$ of analyzed survey papers focused on API evolution in Java libraries.
\cite{https://doi.org/10.1002/smr.328} introduced the terms breaking and non-breaking changes when studying Java API changes. They proposed five strategies for introducing backward-compatible changes. 
Deprecation instead of deletion of functionality allows consumers to use previous versions while marking them as outdated. Delegation forwards outdated method calls to successor methods. Naming conventions, e.g., version numbers in method and class names, help developers to navigate versioned APIs. Runtime switches dynamically load old library versions instead of raising runtime errors. Interface querying allows consumers to request a specific method version via a facade object. 
\cite{10.1145/3447245} conducted interviews and surveys with developers and identified several strategies to reduce or delay consumer-side breaking change impacts.
Providers maintained old interfaces to prolong the transition period for consumers. They released major and minor versions for features and fixes in parallel. Consumers then decided when to migrate to the next major version. As a resulting challenge, providers had to maintain several separate interfaces, called opportunity cost. 
\cite{Wu2016} analyzed Java API changes, and they and \cite{10.1145/3447245} recommended consumers to encapsulate external API dependencies within a facade object to reduce the dependencies and, hence, the change impact.

\cite{6649592} discovered that REST APIs are more change-prone than Java APIs after conducting an empirical study. Further, they identified $6$ additional challenges for REST API evolution compared to source code evolution, e.g., deleted methods are not recoverable since previous versions are not accessible after shutdown and REST APIs require authorization over the network.
\cite{10.1145/3470133} concluded their literature survey with the need for web API evolution research.
\cite{KARABEYAKSAKALLI2021111014} identified synchronous communication with REST and event-driven publish-subscribe communication as the most preferred communication patterns in MSA. They concluded that both approaches are oftentimes combined: REST APIs publish explicit interfaces to the outside world, while publish-subscribe communication loosely couples the internal service architecture for state-changing operations.

\cite{zimmermann_et_al:OASIcs:2020:11826} introduced the microservice API pattern (MAP) framework for API design and evolution. 
They proposed five evolution patterns: running multiple versions in parallel, limited and unlimited lifetime guarantees for backward compatibility, experimental previews without promising stability, and aggressive obsolescence, i.e., shutting down previous versions with a fixed date~\citep{10.1145/3361149.3361164}. 
\cite{ESPINHA201527} interviewed $6$ developers who criticized that early API versions are unstable and change regularly without notice. They formulated basic recommendations, such as staying backward-compatible, exposing some stability status information, and monitoring the system to identify the consumers per feature.
Similarly, \cite{Wu2016} recommended Java API providers should publish their API stability expectations.
\cite{7196531} investigated REST API evolution strategies and found both strategies of running single and multiple concurrent versions in practice. They identified the disconnected source code, documentation, and change log artifacts as challenging because developers must manually link all artifacts to identify changes and their impact. The authors recommended generating customized change logs per consumer.
\cite{8385157} analyzed 500 REST APIs and found that almost half of them automatically generate documentation, e.g., with Swagger UI. The other half provides textual information in varying granularity, complicating automated change identification and impact analysis.

\cite{doi:10.1080/08874417.2018.1520056} investigated microservice opportunities and challenges by interviewing $19$ practitioners. Amongst other organizational challenges, they reported that API changes require intensive communication and coordination. Further, very few practitioners had experience with event-driven communication.
\cite{https://doi.org/10.1002/spe.2967} conducted a grey literature review for microservice API technologies and concerns.
They identified three concerns regarding API gateway design, API versioning, and API testing and test case generation. 
\cite{9825762} analyzed microservice-related StackOverflow questions and identified technical communication as the major challenge in the service construction phase. In the governance phase, they identified further challenges, including defining API standards and API gateway design. 
The proposed solution strategies for these challenges included utilizing event-driven communication and creating GitHub examples for API usage scenarios.
\cite{DETOLEDO2021110968} interviewed $25$ practitioners about architectural technical debts in microservice systems. Interview partners acknowledged that poor REST API design results in API instability, regular breaking changes, and increased difficulty of maintaining backward compatibility. The authors proposed the API-first approach as a solution, which also reduces the tight coupling between services.
\cite{8703917} interviewed $13$ companies and found that inappropriate service boundaries lead to tight coupling and regular changes. The multiple parallel versions then increased the debugging complexity.
Similarly, \cite{app11177856} reported that the system design has a direct impact on microservice evolution and proposed detecting antipatterns in source code and identifying technical debt as future research directions.

In summary, related works reported on the strategies for backward compatibility and versioning to support outdated consumers and recommended the API-first approach. Regarding the challenges, they reported on communication and coordination overheads when migrating API changes, the importance of the API design's quality, and increased code complexity and technical debt when supporting multiple versions.
However, they did not present and discuss the underlying reasons for the challenges or how to solve them sustainably. 
In this work, we formulated a comprehensive list of strategies and challenges for microservice API evolution actively used in practice.
We discovered that close communication and collaboration between teams is not only a well-known challenge but an actively performed and expected strategy in MSA. We identified that the collaboration strategy results from the manual change impact analysis challenge and leads to the problem of tight organizational coupling.
Further, we discovered that the established strategies for compatibility and versioning create a new problem of consumer lock-in. In turn, the consumer lock-in degrades the API design and increases technical debt without any resolution strategy.
To the best of our knowledge, we are the first to formulate such a comprehensive list, to define the problems of tight organizational coupling and consumer lock-in including their underlying challenges, and to propose open research directions addressing them.

\section{Conclusion}
\label{sec:conclusion}

The API evolution process in MSA suffers from the loose coupling between microservices and leads to communication overheads and backward compatibility necessity.
In this work, we conducted semi-structured interviews with $17$ developers, architects, and managers in $11$ companies and reported their strategies and challenges for API evolution. 

In summary, we discovered six strategies and six challenges for REST and event-driven communication techniques. 
The strategies mainly focus on API backward compatibility, versioning, and close collaboration between teams when introducing breaking changes. 
The challenges illuminate the manual change impact analysis efforts, ineffective communication of changes, and consumer reliance on outdated API versions. 
From our findings, we formulated relationships between strategies and challenges and discovered two problems in the microservice API evolution process. Tight organizational coupling undermines the loose technical coupling of microservices by regularly requiring communication and collaboration between development teams. Consumer lock-in increases technical debt and degrades the API design over time by enforcing continuous support for outdated API versions without a clear resolution strategy. We proposed two relevant research directions to mitigate these two problems. 

Future work includes studying the evolution of event-driven communication in particular, which many participants disregarded during our study. Furthermore, based on our insights, we propose a new study that investigates the two problems in MSA API evolution and evaluates approaches to mitigate them.

\bibliographystyle{elsarticle-harv} 
\bibliography{cas-refs}

\end{document}